\documentclass[twocolumn,aps,prb,footinbib,floatfix,superscriptaddress]{revtex4-2} 
\usepackage[utf8]{inputenc}

\usepackage{amsmath,amssymb,bm}
\usepackage{xcolor,graphicx,siunitx}
\definecolor{dodgerblue}{HTML}{1E90FF}
\definecolor{columbiablue}{HTML}{87AFC7}
 
\usepackage[colorlinks=true,citecolor=dodgerblue,linkcolor=dodgerblue,urlcolor=dodgerblue]{hyperref}

\usepackage{tikzsymbols}
\usepackage{slashed}

\newcommand{\braOket}[3]{\langle #1|#2|#3\rangle}
\renewcommand{\vec}[1]{\boldsymbol{#1}}

\makeatletter
\def\maketitle{
\@author@finish
\title@column\titleblock@produce
\suppressfloats[t]}
\makeatother

\begin{document}
\newcommand{\thetitle}{Cyclic structure of Landau levels in transition metal dichalcogenide semiconductors}
\title{\thetitle}
\author{Peize Ding}
\email[]{pd2714@columbia.edu}
\affiliation{Department of Physics, Columbia University, New York, NY, 10027, USA}
\author{Nishchhal Verma}
\affiliation{Department of Physics, Columbia University, New York, NY, 10027, USA}
\author{Raquel Queiroz}
\email[]{raquel.queiroz@columbia.edu}
\affiliation{Department of Physics, Columbia University, New York, NY, 10027, USA}
\affiliation{Center for Computational Quantum Physics, Flatiron Institute, New York, New York 10010, USA}
\date{\today}

\begin{abstract}
Transition metal dichalcogenides (TMDs) exhibit unconventional Landau level (LL) spectra that cannot be fully captured by an effective mass approximation or a minimal two-band Dirac model. Namely, TMDs show an anomalous, upward-sloping zeroth LL in the valence band and an asymmetric orbital magnetization between electron and hole bands. In this paper, we employ a continuum three-band model to derive analytic constraints on the LL spectrum of the $K$ and $K'$ valleys at weak magnetic fields. This model highlights the cyclic structure of the LL spectrum inherited from $C_3$ symmetry, providing both analytical tractability and an accurate description of the band geometry in the low energy approximation of the valleys.
We compare our results against numerical calculations using the three-band tight-binding model of Ref.\cite{Liu2013} and a distorted kagomé lattice model. We find that the Landau levels of the $K$ and $K'$ valleys show a cyclic structure which explains their anomalous slope and magnetization asymmetry. This asymmetry can be traced to the topological obstruction of TMD semiconductors.
We further analyze the impact of disorder, finding that the zeroth LL exhibits partial robustness against certain off-diagonal perturbations, in contrast to the exact index-theorem protection of massive Dirac particles. 
Our results establish a direct link between orbital structure, band topology, and magnetic response in TMDs.
\end{abstract}

\maketitle

\section{Introduction}

Transition metal dichalcogenides (TMDs) are layered materials with rich electronic \cite{mak2010atomically, tongay2012thermally, ross2013electrical, radisavljevic2011single, lembke2012breakdown, lin2012mobility, bao2013high, larentis2012field} and optical properties \cite{splendiani2010emerging,zeng2013optical, koperski2017optical}. Monolayer TMDs in the $1$H phase crystallize in a trigonal prismatic structure where the transition metal (Mo, W) is sandwiched between two layers of chalcogen atoms (S, Se, Te). The low-energy electronic structure near the K and K$'$ valleys of the Brillouin zone is predominantly derived from the transition metal $d$-orbitals, specifically the $d_{z^2}$, $d_{x^2 - y^2}$, and $d_{xy}$ orbitals \cite{Liu2013}. In many TMDs, these orbitals open a direct band gap at the $K$ and $K'$ points. Their lack of inversion symmetry and strong spin-orbit coupling leads to a spin splitting in the valence bands, making these materials ideal candidates for valleytronics and spintronic applications \cite{zibouche2014transition, schaibley2016valleytronics, lin2019two, yang2024valleytronics, cao2011mos_2, xiao2012coupled, lee2017valley}.

A widely used simplification for modeling carriers in these systems is the effective mass approximation that treats the conduction and valence bands as approximate parabolic bands at low energies, characterized by an effective mass \cite{kormanyos2015k, he2023electronic, kumar2012electronic}. Despite the simplicity, the model is successful in capturing many electronic properties of TMDs, ranging from SdH oscillations \cite{fallahazad2016shubnikov} to effective moiré potentials in twisted TMD bilayers \cite{wu2019topological, morales2024magic, smolenski2016tuning}. However,
recent advances in electronic structure~\cite{xiao2010berry} have highlighted the limitations of the effective mass approximation, which fails to properly account for the quantum geometry of electronic wavefunctions. These geometric aspects play a central role in determining optical selection rules~\cite{zeng2013optical, koperski2017optical, wang2017valley, scrace2015magnetoluminescence}, optical spectral weight~\cite{onishi2024universal, verma2024instantaneous, kruchkov2023spectral}, exciton binding energies~\cite{yu2015valley, wang2018colloquium, berkelbach2013theory, robert2017fine}, exciton stiffness~\cite{rossi2021quantum, hu2022quantum, hu2023effect, verma2024geometric}, and anomalous Hall effects~\cite{qian2014quantum, kang2019nonlinear}, among other phenomena. A minimal two-band Dirac model is often used to incorporate these effects~\cite{kormanyos2015k, xiao2007valley, yao2008valley, crepel2024chiral}.
This gapped Dirac model, however, has fundamental limitations that are the most apparent in the Landau level (LL) spectrum as we explore in the present work.

Contrary to expectations for hole-like bands, the energy of the zeroth LL in the valence band of TMDs \emph{increases} with magnetic field~\cite{kormanyos2015k, wang2015magneto, wang2017valley, xuan2020valley, srivastava2015valley, macneill2015breaking, liu2020landau}. Since the slope of LL vs magnetic field has contributions both from the effective mass and from orbital magnetization, it has been argued that orbital magnetization of a two-band model can explain this anomalous LL~\cite{wang2015magneto, xuan2020valley}; however, in any such framework, the orbital magnetization contributes equally to both conduction and valence bands, regardless of the dispersion of each band, as detailed in Appendix~\hyperref[app:kdotp]{A}. Including a third orbital is therefore unavoidable for an accurate description of the LLs of TMDs, balancing analytical tractability and a faithful representation of the band geometry. 
 
Motivated by this observation, we develop a continuum three-band model whose structure is constrained by \(C_3\) symmetry. We show that this model naturally hosts a \emph{cyclic} structure that dictates the LL mixing rules, providing an analytical explanation for the anomalous zeroth LL slope and magnetization asymmetry. To illuminate the physical origin of this behavior, we map the TMD model onto a distorted kagome lattice, establishing a direct link between the anomalous LLs and the material's underlying topological obstruction. Finally, we explore the consequences of this unique cyclic LL structure, demonstrating that it confers a partial, selection-rule-based robustness against specific types of disorder, in stark contrast to the index-protected zero modes of Dirac systems.

The paper is organized as follows. In Sec.~\ref{sec:C3}, we analyze the general constraints imposed by $C_3$ symmetry on the LL energies and wavefunctions in three-band models with symmetry-related orbitals. In Sec.~\ref{sec:butterfly}, we numerically compute the LL spectra of TMDs using both a three-band tight-binding model and a distorted kagome lattice model. The remainder of the paper, Sec.~\ref{sec:disorder}, is devoted to studying the impact of disorder on the LLs in TMDs, highlighting the remarkable robustness of the zeroth LL against certain perturbations, before summarizing our findings and their implications in Sec.~\ref{sec:conclusion}.

\section{Constraints on LLs from $C_3$ symmetry}\label{sec:C3}

Semiconducting TMDs break inversion symmetry while preserving time-reversal and $C_3$ rotational symmetry~\cite{Liu2013}. A tight-binding fit to the \textit{ab initio} band structure that respects the symmetries requires at least three orbitals~\cite{Liu2013}, which transform as $C_3$ eigenstates with eigenvalues $\omega^0$, $\omega^1$, and $\omega^2$, where $\omega = \exp(2\pi i / 3)$. {We label the orbitals by $\gamma = 0, 1, 2$, based on their $C_3$ eigenvalues $\omega^\gamma$}. Physically, they correspond to the metal-derived $d_{z^2}$ and $d \pm i d$ orbitals, {mixed with chalcogen-derived $p$-orbitals}. Because of the underlying $C_3$ symmetry, it is convenient to introduce the \emph{clock} matrix $Z$ and the \emph{shift} matrix $X$, defined as  
\begin{equation}\label{eq:clock_shift}
Z = 
\begin{bmatrix}
\omega^0 & 0 & 0 \\
0 & \omega^1 & 0 \\
0 & 0 & \omega^2
\end{bmatrix},
\qquad
X = 
\begin{bmatrix}
0 & 0 & 1 \\
1 & 0 & 0 \\
0 & 1 & 0
\end{bmatrix},
\end{equation}
in the orbital basis $\left(|0\rangle, |1\rangle, |2\rangle\right) \equiv (|\gamma\rangle,\; \gamma=0,1,2)$. 
Here $Z$ represents the action of the $C_3$ rotation, while $X$ cyclically permutes the orbitals. 
Together they generate the algebra of the three-state clock model, satisfying the relation
\begin{equation}
    X Z = \omega Z X,
\end{equation}
which encodes the cyclic structure imposed by $C_3$ symmetry. Although the Hamiltonian is not invariant under the shift operator $X$, the family of Hamiltonians is closed under its action. We say that the Hamiltonian family is \emph{cyclic} if it is closed under the action of $X$,
\begin{equation}
    XH(\vec{k}; \epsilon_i, \alpha_i, v_{ij})X^{-1} = H(\vec{k}; \epsilon_{i-1}, \alpha_{i-1}, v_{i-1,j-1}),
\end{equation}
with all indices understood modulo 3 (see Appendix~\hyperref[app:cyclicity]{B} for further discussion).

We focus on TMDs where the valence band maximum and conduction band minimum lie at the $K$ point, including MoTe$_2$, MoSe$_2$, WSe$_2$, and WTe$_2$. $C_3$ symmetry dictates that $h_{\gamma \gamma'}(C_3\vec{k}) = \omega^{-\gamma + \gamma'} h_{\gamma \gamma'}(\vec{k})$. To lowest order in momentum $\vec{k}$ (quadratic for diagonal elements and linear for off-diagonal elements), this symmetry constraint determines the structure of the Hamiltonian.
Consequently, the Hamiltonian expanded around the $K$ point takes the form
\begin{align}\label{eq:C3_Hk}
H(\vec{k}) \approx 
\left[\begin{array}{ccc}
\epsilon_{0} + \alpha_{0}k^2 & 
v_{01} k_+ & 
v_{02} k_- \\
v_{10} k_- &
\epsilon_{1} + \alpha_{1}k^2 &
v_{12} k_+ \\
v_{20} k_+ &
v_{21} k_- &
\epsilon_{2} + \alpha_{2}k^2
\end{array}\right],
\end{align}
in the basis of $(|0\rangle, |1\rangle, |2\rangle)$ orbitals,
where $\vec{k}$ is defined relative to the $K$ point, $k^2 = k_x^2 + k_y^2$, $k_{\pm} = k_x \pm i k_y$ ~\cite{wang2015magneto}.
   Here, $\epsilon_\gamma$ denotes the on-site energy and $\alpha_\gamma$ captures the intra-orbital hopping via a quadratic expansion in momentum. 
   The inter-orbital coupling is described by $v_{\gamma \gamma'}$, with Hermiticity requiring $v_{\gamma \gamma'} = v_{\gamma' \gamma}^*$. For the other valley $K'$, time-reversal symmetry imposes that $H(\vec{K}'+ \vec{k}) = [H(\vec{K} - \vec{k})]^*$. Because of the constraints of $C_3$ symmetry, there is a cyclic structure associated to the Hamiltonian, which we discuss in detail in Appendix~\hyperref[app:cyclicity]{B}.

We next study the LL structure of this $C_3$ symmetric Hamiltonian. We focus on the case with small magnetic field, such that the magnetic length $l_B = \sqrt{\hbar/eB}$ is much larger than the lattice constant $l_B \gg a_0$. We further set $\hbar = e = a_0 =  1$ so that $l_B^{-2} = B$.
{In the presence of perpendicular magnetic field, the momentum $\vec{k}$ should be replaced by the gauge-invariant canonical momentum $\vec{\Pi} = \vec{k} + \vec{A}$ \cite{bernevig2013topological}, where $\vec{A}$ is the vector potential.}
The LLs around the $K$ ($K'$) point can then be calculated through the continuum model approximation \cite{bernevig2013topological,he2023electronic}
\begin{equation}\label{eq:LL_operators}
\begin{split}
k_+ \rightarrow  \sqrt2 \hat a / l_B,\  k_- \rightarrow \sqrt2 \hat a^\dagger / l_B, \  k^2 \rightarrow 2 (\hat a^\dagger \hat a + 1 / 2) / l_B^2.
\end{split}
\end{equation}
Here the {dimensionless ladder operators $\hat a = (l_B / \sqrt2)\Pi_- $ and $\hat a^\dagger =  (l_B / \sqrt2)\Pi_+$, where $\Pi_{\pm} = \Pi_x \pm i \Pi_y$, which satisfies $[\hat a, \hat a^\dagger] = 1$. As in one-dimensional harmonic oscillator,
\begin{equation}
\hat{a}|n\rangle=\sqrt{n}|n-1\rangle, \quad \hat{a}^{\dagger}|n\rangle=\sqrt{n+1}|n+1\rangle,
\end{equation}
where $|n\rangle$ denotes the $n$-th LL and $n \geq 0$ is an integer.}
We see the Hamiltonian has the structure
\begin{widetext}
\begin{equation}\label{eq:C3_in_B}
H = \left[\begin{array}{ccc}
\epsilon_{0} + 2B\alpha_{0}(\hat a^\dagger \hat a  + \frac{1}{2}) & 
\sqrt{2B} v_{01} \hat a  & 
\sqrt{2B} v_{02} \hat a^\dagger \\
\sqrt{2B} v_{10} \hat a^\dagger &
\epsilon_{1} + 2B\alpha_{1}(\hat a^\dagger \hat a  + \frac{1}{2}) &
\sqrt{2B} v_{12} \hat a \\
\sqrt{2B} v_{20} \hat a &
\sqrt{2B} v_{21} \hat a^\dagger &
\epsilon_{2} + 2B\alpha_{2}(\hat a^\dagger \hat a  + \frac{1}{2})
\end{array}\right]    .
\end{equation}
\end{widetext}

To obtain the energy dispersion of the LLs at small magnetic field $B$, we apply second-order perturbation theory using the off-diagonal matrix elements $\sqrt{2B} \sum_{\gamma} v_{\gamma, \gamma+1}  + \text{h.c.}$ as the perturbation, where $\gamma \pm 1$ is understood under mod $3$. This approach is justified because the perturbation strength is controlled by the small parameter $\sqrt{B}$. The unperturbed states are the bare Landau levels $|n,\gamma\rangle$, labeled by the LL index $n$ and orbital character $\gamma$.
We find that the linear order correction to energies take the form
\begin{equation}\label{eq:LL_perturbation_energy}
\begin{split}
    \epsilon_{n,\gamma}(B) = \epsilon_\gamma + \left({\alpha_\gamma} + 2\mu_{\gamma, \gamma + 1}\right)B + \\
    \quad 2(\alpha_\gamma+\mu_{\gamma, \gamma + 1} + \mu_{\gamma, \gamma  - 1})nB
\end{split}
\end{equation}
with $n$ is the LL index, and $\mu_{\gamma\gamma^\prime} \equiv |v_{\gamma \gamma^\prime}|^2 / (\epsilon_\gamma - \epsilon_{\gamma^\prime})$ arises from perturbation theory.
Higher-order corrections can be obtained by expanding the Hamiltonian to higher order in $\vec{k}$ within the continuum model approximation. 
From Eq.~\eqref{eq:LL_perturbation_energy}, we can extract the slope of the zeroth LL with respect to the magnetic field:
\begin{equation}\label{eq:LL_slope}
    s^0_\gamma \equiv \left.\frac{d \epsilon_{n, \gamma}(B)}{dB}\right|_{B = 0} = \alpha_\gamma + 2\mu_{\gamma, \gamma + 1},
\end{equation}

The slope of the zeroth LL $s_\gamma^0$, comprises two distinct physical contributions: a conventional term $\alpha_\gamma$ related to the intra-orbital band curvature, and a quantum geometric term $2 \mu_{\gamma, \gamma+1}$ arising from magnetic-field-induced coupling to other orbitals, which manifests as an orbital magnetization, as derived in Appendix~\hyperref[app:kdotp]{A}. 
   
Notably, to lowest order in $B$, the zeroth LL exhibits a unique mixing behavior: it couples only to the $\gamma + 1$ orbital, with no contribution from $\gamma - 1$. This selective coupling is a direct consequence of the cyclic structure of the Hamiltonian and stands in sharp contrast to higher LLs ($n > 0$ in Eq.~\eqref{eq:LL_perturbation_wf}), which generically have support on all three orbitals.

We apply perturbation theory to analyze the structure of the wavefunctions, revealing how different orbitals hybridize under a magnetic field. The perturbed eigenstate to linear order in $\sqrt{B}$ is
\begin{equation}\label{eq:LL_perturbation_wf}
\begin{split}
    |\psi_{n, \gamma}\rangle = |n, \gamma\rangle 
    + \frac{v_{\gamma, \gamma + 1} \sqrt{2B(n + 1)}}{\epsilon_{\gamma}- \epsilon_{\gamma + 1}} |n + 1, \gamma+1\rangle 
    +\\ \frac{v_{\gamma, \gamma - 1} \sqrt{2Bn}}{\epsilon_{\gamma}- \epsilon_{\gamma - 1}} |n - 1, \gamma - 1\rangle,
\end{split}
\end{equation}
up to an overall normalization factor, where $|n, \gamma\rangle$ denotes the $n$-th Landau level on orbital $\gamma$. Particularly,
 the zeroth LL ($n=0$) with $\gamma=1$ (the orbital character of the valence band at the $K$ valley) becomes
\begin{equation}
    |\psi^{K}_{0,1}\rangle = |0,1\rangle 
    + \sqrt{2B}(\mu_{12}/v_{12}^*)\, |1,2\rangle.
\end{equation}
Thus, the valence-band zeroth LL in the $K$ valley mixes only with the $n=1$ LL or orbital 2. On the other hand, the $n=1$ state already involves all three $C_3$ orbitals.

This discussion highlights that there is an intrinsic cyclic structure associated with the LL wavefunctions, inherited from the Hamiltonian.
To characterize this cyclicity, we define the operators
\begin{equation}
    \hat{X}_{\circlearrowleft} = \hat a X, 
    \qquad 
    \hat{X}_{\circlearrowright} = \hat a X^\dagger,
\end{equation}
where $\hat a$ is the Landau-level lowering operator (Eq.~\eqref{eq:LL_operators}) and $X$ is the orbital shift introduced in Eq.~\eqref{eq:clock_shift}. 
For the LL wavefunctions of the $K$ ($K'$) valley in a positive $z$-directed magnetic field (Eq.~\eqref{eq:LL_perturbation_wf}), we find that the expectation value $\langle\hat{X}_{\circlearrowright}\rangle \neq 0$  ($\langle\hat{X}_{\circlearrowleft}\rangle \neq 0$), while  $\langle\hat{X}_{\circlearrowleft}\rangle = 0$ ($\langle\hat{X}_{\circlearrowright}\rangle = 0$). 
This valley-contrasting selection rule implies that circularly polarized light of a given chirality couples selectively to LLs in only one valley~\cite{rose2013spin, wang2017valley}, directly linking the cyclic orbital structure to valley-dependent optical activity. Having established these analytical constraints, we next turn to numerical calculations on realistic lattice models to verify these predictions and explore their connection to the global band topology.

\section{Landau levels of TMDs}\label{sec:butterfly}

The $C_3$ symmetry analysis in Sec.~\ref{sec:C3} establishes the algebraic structure that governs the form of the Landau level wavefunctions. 
To assess how these constraints manifest in realistic band structures, we now turn to explicit calculations of the LL spectra using a three-band tight-binding model. 
This also enables us to explore the topological origin of the anomalous zeroth LL dispersion by mapping the problem to a distorted kagome lattice.

\subsection{Tight-binding model}

\begin{figure}[tbp]
		\centering
		\includegraphics[width=0.48\textwidth] {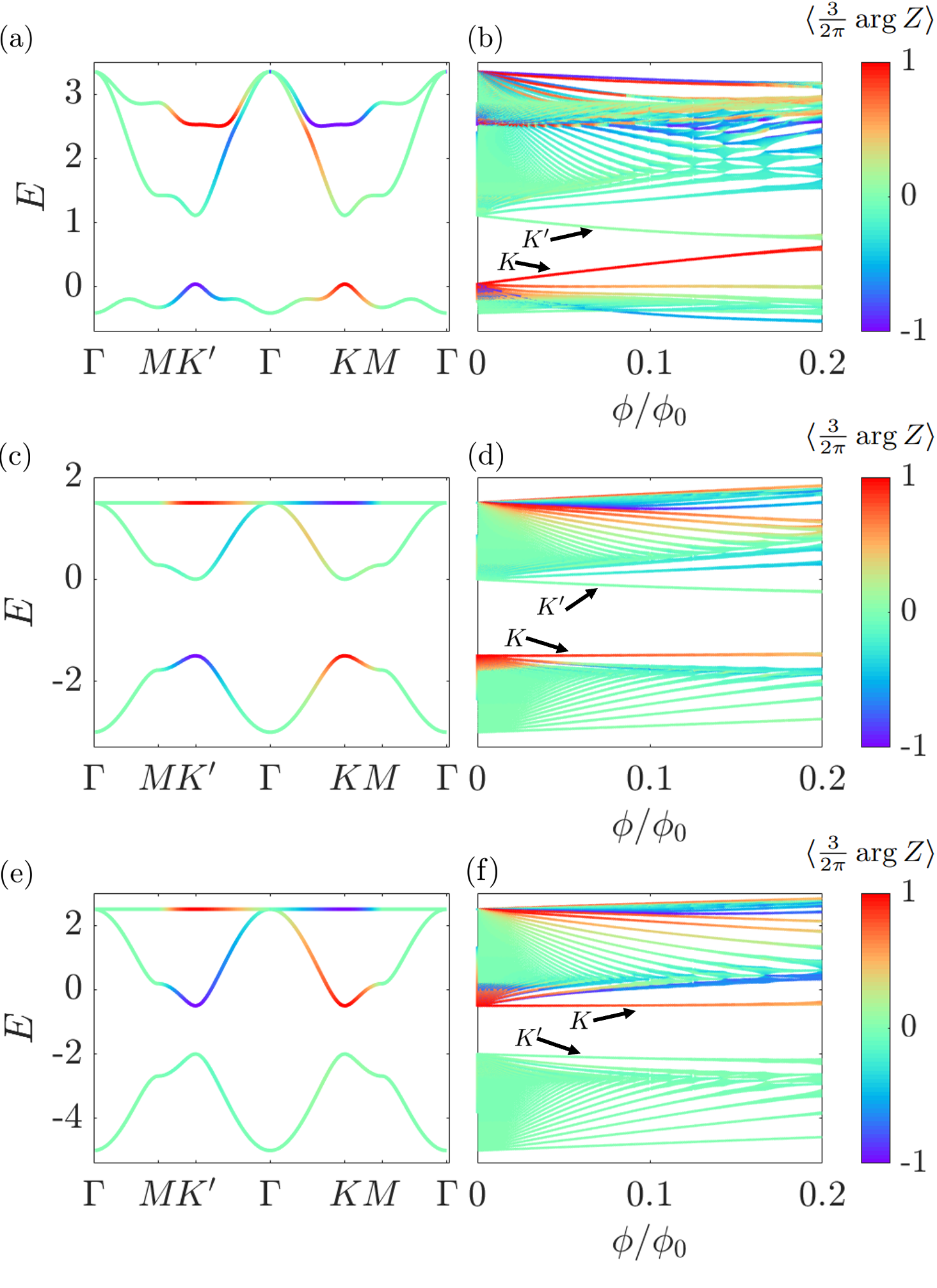}
		\caption{(a) Band structure for the three-band tight-binding model of MoTe$_2$, and (b) its LL spectrum as a function of magnetic flux $\phi$ in units of the flux quantum $\phi_0 = h / e$. The energy unit is 1 eV.
			(c), (e) Band structures and (d), (f) LL spectra of the distorted kagome lattice for $t = 0.5, t^\prime = 1$ and $t = 1.5, t^\prime = 1$, respectively, as functions of magnetic flux $\phi$ in units of the flux quantum. The color denotes the expectation value of $S = (3/2\pi) \arg Z$, where $S=0, 1, -1$ correspond to the $\omega^0 (\gamma=0), \omega^1 (\gamma=1)$, and $\omega^2 (\gamma=2)$ orbital characters, respectively. $K$ and $K'$ in the LL plots indicate which valley the lowest (highest) LL at the conduction (valence) band comes from. 
			}
		\label{fig:band_butterfly}
\end{figure}

A three-band tight-binding model, fitted to the ab initio band structure \cite{Liu2013}, is detailed in Appendix~\hyperref[app:model]{C}. We neglect spin-orbit coupling to focus on orbital effects. The MoTe$_2$ band structure is shown in Fig.~\ref{fig:band_butterfly}(a). To quantify orbital character, we can calculate the expectation value of $S \equiv ({3}/{2\pi}) \operatorname{arg} Z$,
whose eigenvalues $0, 1, -1$ correspond to the orbital characters $\omega^0, \omega^1, \omega^2$, respectively. $S$ is related to the clock matrix via $Z = e^{2\pi i S/3}$.
In Fig.~\ref{fig:band_butterfly}, color indicates the expectation value of \( S \). At the high-symmetry \( C_3 \) points (\( \Gamma, K, K' \)), the wavefunctions are eigenstates of \( S \) due to the $C_3$ symmetry. The valence band exhibits orbital characters \( \omega^0, \omega^1, \omega^2 \) at \( \Gamma, K, K' \), respectively, a universal feature of TMDs~\cite{Liu2013}. This structure reflects the obstructed atomic limit of TMD semiconductors, where the charge center lies at the center of the triangle (Wyckoff position \( b \); see Fig.~\ref{fig:tri_kagome}(a)), misaligned with the chalcogen sites~\cite{holbrook2024real, cano2022topology, jung2022hidden}. To leading order, the maximally localized Wannier function is centered on the elementary triangles (Fig.~\ref{fig:tri_kagome}(a)) and has orbital support on the triangle vertices
\begin{equation}\label{eq:Wannier_TMD}
    |\psi_j\rangle = \frac{1}{\sqrt{3}} \left(|0\rangle + \omega^j|1\rangle + \omega^{-j}|2\rangle\right), \quad j = 0, 1, 2.
\end{equation}
This function, analogous to $sp^2$ hybridization of $d$-orbitals, preserves $C_3$ symmetry and underlies the orbital characters observed at high-symmetry points~\cite{eck2023recipe, jung2022hidden, qian2022c}.

We now turn to the complete LL spectrum of the model using the Hofstadter spectrum. To isolate the inter-atomic orbital effects, we neglect the coupling between spin, atomic angular momentum, and the magnetic field (see~\cite{xuan2020valley} for further discussion). 
To compute the LL energies on the lattice, we apply the Peierls substitution to the hopping terms: for a hopping from site \( \vec{r}_i \) to \( \vec{r}_j \), the matrix element is modified as
\begin{equation}
    t_{ij} \rightarrow t_{ij} \exp\left(\dfrac{i e}{\hbar }\int_{\vec{r}_i}^{\vec{r}_j} \vec{A} \cdot d\vec{l} \right)
\end{equation}
where \( \vec{A} \) is the vector potential~\cite{hofstadter1976energy, bernevig2013topological}. 

The resulting LL spectrum is shown in Fig.~\ref{fig:band_butterfly}(b). Notably, the zeroth LL at the valence (conduction) band edge exhibits a positive (negative) slope, opposite in sign to the effective mass at the \( K \) (\( K' \)) point. This anomalous dispersion is a generic feature across TMDs, as detailed in Appendix~\hyperref[app:butterfly]{D}, and has been observed experimentally in optical measurements~\cite{wang2017valley}.

We note that the lattice calculation is consistent with the perturbation theory arguments presented in the previous section (see Appendix~\hyperref[app:model]{C}). As shown in Fig.~\ref{fig:band_butterfly}(a), the valence band at the $K$ point has orbital character $S=-1$ ($\gamma=2$), and the slope of its zeroth LL is $s_2^0 = \alpha_2 + 2\mu_{20}$. The positive parameters $\alpha_2$ and $\mu_{20}$ (derived from Ref.~\cite{Liu2013}) explain the anomalous upward slope observed numerically and experimentally.

\subsection{Orbital magnetization and $g$-factor}

\begin{table*}[t]
\label{table:g-factor}
\centering
\caption{Orbital contribution to $g$-factors of conduction and valence band with spin up from experiment and theory for WSe$_2$.}
\begin{tabular}{lccccc}
\hline\hline
 & \begin{tabular}{c}Tight-binding\\(this work)\end{tabular}
 & \begin{tabular}{c}Measurements\\Robert et al.~\cite{robert2021measurement}\end{tabular}
 & \begin{tabular}{c}Calculations\\Xuan et al.~\cite{xuan2020valley}\end{tabular}
 & \begin{tabular}{c}Calculations\\Deilmann et al.~\cite{deilmann2020ab}\end{tabular}
 \\
\hline
$g_{\textrm{orb}, c}$ & -3.89 & -2.84 & -2.91 & -2.97 \\
$g_{\textrm{orb}, v}$   & -4.26 & -5.10 & -4.81 & -4.91 \\
\hline\hline
\end{tabular}
\end{table*}

A notable feature revealed by magneto-optical measurements on monolayer TMDs is the anomalously large effective Land\'e $g$-factor, especially in the valence band. In WSe$_2$, for example, the valence-band $g$-factor has been measured to reach $g_v \sim 6$~\cite{robert2021measurement}, far exceeding the free-electron value $g_0 \approx 1$ (the spin $g$-factor under the convention without the $1/2$ factor). The total effective $g$-factor receives contributions from several sources: the intrinsic electron spin, the valley magnetic moment, the atomic orbital angular momentum, and additional many-body renormalizations~\cite{xuan2020valley}.

The theoretical framework allows us to estimate the orbital magnetic moment contribution to the effective $g$-factor~\cite{xuan2020valley}. As derived in Appendix~\hyperref[app:kdotp]{A}, the orbital $g$-factor is given by
\[
g_{\text{orb},\gamma}
= \mu_{\gamma,\gamma-1} - \mu_{\gamma,\gamma+1},
\]
where we set $\mu_B = 1$.
Crucially, this formula implies that at least three bands are required to generate the experimentally observed asymmetric orbital magnetization between conduction and valence bands (see Appendix~\hyperref[app:kdotp]{A}).

Using continuum-model parameters derived from the tight-binding model, we can estimate these values. Consistent with Fig.~\ref{fig:band_butterfly}(a), we identify the conduction band at the $K$ point with orbital index $\gamma=0$ and the valence band with $\gamma=1$. For WSe$_2$, the calculation yields
\[
g_{\text{orb},c} = \mu_{02} - \mu_{01} \approx -4.26,
\quad
g_{\text{orb},v} = \mu_{10} - \mu_{12} \approx -3.89 .
\]
The values for the opposite valley have the opposite sign due to time-reversal symmetry.

We compare these tight-binding estimates with experimental measurements and other ab-initio theoretical results in Table~\ref{table:g-factor}. The experimental value includes the spin Zeeman contribution, which we subtract in order to isolate the orbital response for the spin-up species. From this comparison, we find that the tight-binding-derived orbital moment accounts for more than $80\%$ of the measured valence-band orbital $g$-factor, even without including atomic orbital contributions or spin-orbit coupling~\cite{xuan2020valley}. This indicates that the dominant source of the orbital magnetization originates from the quantum geometric effects captured in the model we consider.

\subsection{Distorted Kagome lattice}

\begin{figure}[htbp]
		\centering
		\includegraphics[width=0.46\textwidth] {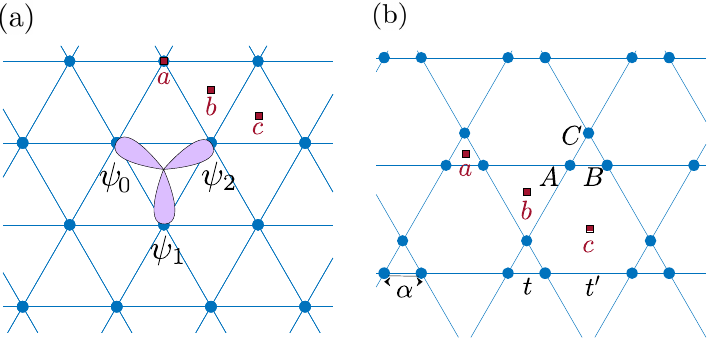}
\caption{(a) Triangular lattice for the TMD model. The lobes demonstrate the Wannier function of the valence band to the lowest order approximation (Eq.~\eqref{eq:Wannier_TMD}). $\hat a, b, c$ labels the three inequivalent Wyckoff positions. Note that the chalcogen atoms are located at Wyckoff positions $c$, which misaligns with the Wannier center.
        (b) Distorted kagome lattice with $\alpha$ controlling the size of the unit cell comparing to the lattice spacing. $\hat a, B, C$ labels the three sites in one unit cell. $t$ ($t^\prime$) denotes the nearest neighbor intra-(inter-)cell hopping strength. 
        }
		\label{fig:tri_kagome}
\end{figure}

To isolate the essential mechanism linking the anomalous LL dispersion to band topology, we now introduce a minimal three-band model on a distorted kagome lattice. This model shares the same symmetries as the TMDs but allows us to directly tune the system between a topologically obstructed phase and a trivial atomic insulator, revealing the origin of the anomalous LL behavior. A close correspondence exists between models of TMD and a distorted kagome lattice model, as shown in Fig.~\ref{fig:tri_kagome}(b). The kagome lattice consists of three sites per unit cell (\( A, B, C \)), with a tunable distortion parameter \( \alpha \), defined as the ratio of intra-cell bond length to the (unit) lattice constant.

Exploiting the shared $C_3$ symmetry, a direct mapping can be established between the TMD Wannier basis and the kagome site basis
\begin{equation}
    (|\psi_1\rangle\ |\psi_2\rangle\ |\psi_3\rangle) \Longleftrightarrow (|A\rangle\ |B\rangle\ |C\rangle),
\end{equation}
ensuring identical actions of $C_3$ and time-reversal in both models. Using this mapping and Eq.~\eqref{eq:Wannier_TMD}, the $C_3$ eigenbasis in the kagome model is
\begin{equation}
    |\gamma\rangle = \frac{1}{\sqrt{3}} \left( |A\rangle + \omega^\gamma |B\rangle + \omega^{-\gamma} |C\rangle \right).
\end{equation}

To be more precise, a tight-binding model on the distorted kagome lattice with nearest neighbor hopping is given by
\begin{equation}\label{eq:kagome_TB}
    H^{\operatorname{kagome}} = -t\sum_{\langle ij\rangle}c_i^\dagger c^{}_j - t^\prime\sum_{\langle ij\rangle^\prime}c_i^\dagger c^{}_j + \text{h.c}
\end{equation}
where $ \langle ij\rangle $ and $ \langle ij\rangle' $ denote intra-cell and inter-cell hoppings, respectively, as illustrated in Fig.~\ref{fig:tri_kagome}(b). The model breaks inversion symmetry when $ \alpha \neq 0.5 $ or $ t \neq t' $ (with the lattice constant set to 1), yielding the same symmetry group as the three-band TMD model~\cite{jung2022hidden, Ingham2025unpublished}.

We show the band structure of $ H^{\operatorname{kagome}} $ for $ t < t' $ and $ t > t' $ in Fig.~\ref{fig:band_butterfly}(c) and (e), respectively. Note that the orbital character depends on the choice of embedding in the cell-periodic parts of Bloch wavefunctions; we adopt a gauge where all sublattices are placed at the origin, which most closely matches the TMD setting.

When $ t < t' $, the valence and conduction bands are inverted at the $ K $ and $ K' $ points, mirroring the behavior in TMDs. In this regime, the system realizes an obstructed atomic limit with the charge center at Wyckoff position $ b $ (see Fig.~\ref{fig:tri_kagome}(b))~\cite{geschner2024band}, and the $ C_3 $ representations at $ \Gamma, K, K' $ match those of the TMD model. In contrast, for $ t > t' $, the charge center shifts to Wyckoff position $ a $, coinciding with the atomic sites and thus indicating a topologically trivial atomic limit. As shown in Fig.~\ref{fig:band_butterfly}(e), no band inversion occurs at $ K $ or $ K' $, and the valence band is purely of $ \omega_0 $ character.
In summary, the distorted kagome lattice offers a minimal model for exploring the relationship between Landau level structure and topological obstruction.

We now examine the LL spectra of the distorted kagome lattice. {While the band structure does not depend on $\alpha$, the LL spectra do depend on the sublattice geometries. To emphasize its energetic and symmetry similarity to TMDs, we consider the geometry $ \alpha = 0 $, where the kagome lattice becomes geometrically equivalent to a triangular lattice}, while retaining distinct intra- and inter-cell hopping amplitudes $ t $ and $ t' $. 

In this limit, the model is topologically obstructed only when $ t < t' $, since the Wyckoff position $ a $ coincides with lattice sites. The resulting spectra are shown in Fig.~\ref{fig:band_butterfly}(d) and (f). In the obstructed case (d), both valence and conduction band zeroth Landau levels (LLs) exhibit anomalous dispersion. In contrast, the trivial case (f) with $ t > t' $ shows normal LL behavior, with the zeroth LL confined within the band.

Crucially, this implies that in the distorted kagome model, whenever the bands are inverted at the valleys (indicative of an obstructed atomic limit), the zeroth LL displays anomalous dispersion. This mechanism mirrors the behavior seen in TMDs, establishing a clear link between topological obstruction and anomalous Landau quantization. Exploring the general relation between topological obstruction and Landau level dispersion will be an interesting direction for future study.

\section{Index theorem and disorder-immune Landau level} \label{sec:disorder}

The cyclic orbital structure derived in Sec.~\ref{sec:C3} not only explains the anomalous LL dispersion but also endows the wavefunctions with a unique internal structure. We now investigate the physical consequences of this structure by examining the robustness of the LLs against disorder in the weak-field regime, contrasting their behavior with the well-understood case of gapped graphene.
The latter can be described by a gapped Dirac Hamiltonian, for which the Atiyah–Singer index theorem ensures the robustness of the zeroth LL against perturbations in the off-diagonal channels, including hopping disorder and fluctuations of the magnetic field \cite{atiyah1963index,aharonov1979ground,Tohru2009quantum}. 
In close analogy, we find that for the three-band Hamiltonian of TMDs in the weak-field regime, certain LLs, most notably the zeroth level, remain protected against specific off-diagonal perturbations.

\begin{figure*}[htbp]
		\centering
		\includegraphics[width=0.75
        \textwidth] {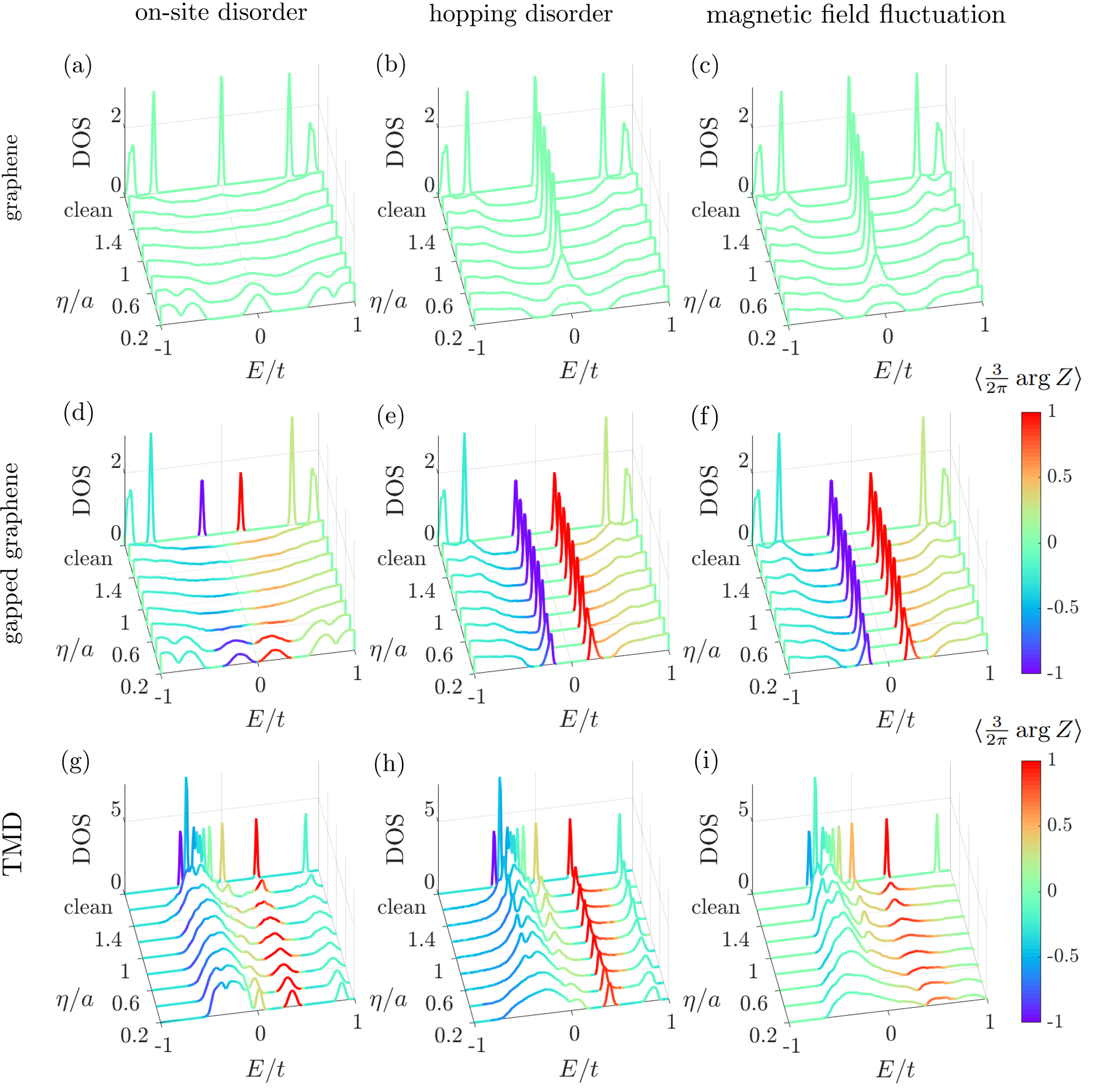}
		\caption{      
        The evolution of density of states (in arbitrary units, scaled for clarity) as a function of energy $E$ as the correlation length $\eta$ increases for graphene with $m = 0, \phi = 0.05\phi_0$ ((a)-(c)), graphene with $m = 0.4 t, \phi = 0.05\phi_0$ ((d)-(e)) and MoTe$_2$ with $\phi = 0.1\phi_0$ ((g)-(i)). For graphene, the color denotes the expectation value of $\sigma_z$. For TMDs, the color denotes the expectation value of $S = \frac{3}{2\pi} \operatorname{arg} Z$. The first, second and third column plots the results with on-site disorder ($\epsilon_i$), hopping disorder ($\delta_{ij}$) and magnetic fluctuation ($\theta_{ij}$), respectively. The density of state for clean samples are also plotted for comparison. For graphene the energy unit $t$ is the nearest neighbor hopping strength. For MoTe$_2$, the disorder strength is $\sigma = 0.1$ eV. The strength of disorder $\sigma = 0.3t$ is used for graphene. $N = 60$ for graphene and $N = 40$ for MoTe$_2$. We averaged over $5$ disorder realizations for a clear qualitative demonstration. This protection is topological in nature, as the number of zero modes of a chiral Dirac Hamiltonian is a topological invariant fixed by the Atiyah-Singer index theorem.}
		\label{fig:three_disorder}
\end{figure*}

\subsection{Disorder-Immune Zeroth LL in Gapped Graphene}

For gapped graphene, the low-energy physics in the $K$ valley is captured by the Dirac Hamiltonian with mass $m$ and Fermi velocity $v$. In this model, the zeroth Landau level (LL) has a wavefunction that is entirely sublattice polarized, $|\psi_0\rangle = (0, |0\rangle)$. Consequently, any off-diagonal perturbation (coupling the two sublattices) strictly vanishes when acting on the zeroth LL, ensuring its robustness. This protection is guaranteed by the Atiyah-Singer index theorem for the Dirac Hamiltonian~\cite{atiyah1963index, aharonov1979ground}.

To illustrate this, we simulate a honeycomb lattice tight-binding model with mass $m$ and spatially correlated disorder (correlation length $\eta$) following Refs.~\cite{PhysRevB.75.235317, Tohru2009quantum}. We provide the details of the model and disorder implementation in Appendix~\hyperref[app:graphene]{E}. We consider on-site, hopping, and phase disorder.

Figures~\ref{fig:three_disorder}(a)–(c) show results for gapless graphene (\( m = 0 \)). For uncorrelated on-site disorder (a), the LL structure is destroyed regardless of \( \eta \). In contrast, hopping (b) and phase (c) disorders in the off-diagonal channels, preserve the zeroth LL when \( \eta \) is large. This robustness arises because short-range disorder couples the \( K \) and \( K' \) valleys, while long-range disorder does not. In the long-range limit, the two valleys are decoupled, and the Atiyah–Singer index theorem protects the zero modes~\cite{aharonov1979ground}.

Figures~\ref{fig:three_disorder}(d)–(f) show the gapped case (\( m \neq 0 \)). Color indicates the expectation value of \( \sigma_z \), reflecting the sublattice polarization. In the clean limit, each valley hosts a zeroth LL at energy \( \pm m \), polarized on opposite sublattices. On-site disorder (d) destroys all LLs, but hopping (e) and phase (f) disorders preserve the zeroth LL for large \( \eta \). Furthermore, since the two zeroth LLs are no longer degenerate, inter-valley scattering is suppressed, enhancing their robustness compared to the \( m = 0 \) case. This is evident from the height of the DOS peaks, which remain sharp and unchanged for \( m \neq 0 \), but are reduced for \( m = 0 \) due to valley mixing.

\subsection{Landau Levels in TMDs}

For TMDs, the low-energy physics in the $K$ valley is described by the Hamiltonian in Eq.~\eqref{eq:C3_in_B}. 
We focus on the lowest LLs originating from the valence and conduction bands, whose approximate wavefunctions are given by Eq.~\eqref{eq:LL_perturbation_wf}:
\begin{equation}
    |\psi_{0,1}\rangle \sim |0,1\rangle + \beta_1 |1,2\rangle, 
    \qquad 
    |\psi_{0,0}\rangle \sim |0,0\rangle + \beta_0 |1,1\rangle,
\end{equation}
where the Landau level state $|n,\gamma\rangle$ denotes the $n$-th LL of orbital $\gamma$, and the mixing coefficients are 
\(\beta_1 = {v_{12}\sqrt{2B}}/({\epsilon_1 - \epsilon_2})\), 
\(\beta_0 = {v_{01}\sqrt{2B}}/({\epsilon_0 - \epsilon_1})\). 

Let us first consider $ |\psi_{0,1}\rangle $. 
Diagonal perturbations do not provide any protection for the robustness of the LLs, similar to the two-band case. 
In contrast, perturbations in the $v_{01}$ channel, defined by
\begin{equation}
    \delta H_{v_{01}} =  
    \sqrt{2B}
    \begin{bmatrix}
        0 & \delta v_{01} a & 0 \\
        \delta v_{01}^* \hat a^\dagger & 0 & 0 \\
        0 & 0 & 0
    \end{bmatrix},
\end{equation}
do not affect $ |\psi_{0,1}\rangle $, since $\delta H_{v_{01}}|\psi_{0,1}\rangle = 0$. 
By contrast, perturbations in the $v_{20}$ and $v_{12}$ channels broaden the LL. 
However, because $\delta H_{v_{20}}|\psi_{0,1}\rangle$ and $\delta H_{v_{12}}|\psi_{0,1}\rangle \propto \sqrt{B}$, the broadening effect is parametrically weak in the small-field regime. 
A similar line of reasoning applies to $ |\psi_{0,0}\rangle $. 

From a symmetry perspective, the $C_3$ rotation symmetry is explicitly broken by the magnetic field in Eq.~\eqref{eq:C3_Hk}, resulting in LLs that are mixtures of $C_3$-symmetric orbitals. 
When the field is weak, the orbital mixing remains small, so the lowest LLs retain partial robustness against off-diagonal perturbations that couple different $C_3$ orbitals. 
For higher LLs, however, the mixing grows with increasing LL index, leading to progressively weaker protection.

We now examine the effect of disorder on LLs in TMDs using the real-space three-band tight-binding model of MoTe$_2$ (see Appendix~\hyperref[app:model]{C}) and introduce disorder in the on-site potential, hopping amplitude, and hopping phase, analogous to the graphene case discussed above.
Figures~\ref{fig:three_disorder}(g)–(i) show the density of states (DOS) under these three disorder types.
As in graphene, on-site disorder (g) broadens all LLs. 

For hopping disorder (h), we focus on off-diagonal terms that couple the $C_3$-symmetric orbitals $d_{z^2}$ and $d \pm i d$, which are particularly relevant for strain-induced perturbations. 
Our numerical simulations reveal that the zeroth LLs of the valence ($\omega^1$) and conduction ($\omega^0$) bands in the $K$ valley retain a degree of robustness: they exhibit sharp spectral peaks even in the presence of disorder, albeit with reduced height when the disorder correlation length is large. 
This behavior highlights the special role of the zeroth LL, whose orbital structure suppresses certain scattering channels. 
In contrast, higher LLs are more strongly broadened, reflecting their greater orbital mixing and reduced protection. 

The situation is qualitatively different for magnetic-field fluctuations, modeled as random phase disorder, as shown in Fig.~\ref{fig:three_disorder}(i). 
Here all LLs are significantly broadened, in stark contrast with the selective robustness seen for hopping disorder. 
The key distinction is that phase disorder introduces perturbations in both diagonal and off-diagonal channels, thereby opening additional scattering pathways that affect all levels. 
As a result, the zeroth LL loses its relative stability and the entire spectrum becomes more diffusive under such perturbations.

This selective protection is not a fundamental topological guarantee as in graphene, but rather an emergent property of the weak-field limit, where the orbital mixing is minimal and dictated by the cyclic selection rules of the Hamiltonian. This robustness is therefore expected to diminish at stronger magnetic fields or for higher Landau levels, where orbital hybridization becomes more complex and involves all three basis states.

\section{Conclusion} \label{sec:conclusion}

This work presents a detailed analysis of Landau level (LL) spectra in TMDs using a symmetry-based three-band model. By examining constraints imposed by $ C_3 $ symmetry in the weak-field limit, we identified characteristic cyclic features in LL energies and wavefunctions. Our numerical simulations revealed an anomalous upward dispersion of the zeroth LL in the valence band, which we explained through an effective mapping to a distorted kagome lattice. This mapping elucidated a connection between the anomalous LL behavior and the topological obstruction inherent in obstructed atomic limits.
We also studied the impact of disorder, showing that the zeroth LL remains particularly robust under off-diagonal hopping disorders. This robustness stems from the cyclic structure of the Hamiltonian and the resulting orbital composition of the wavefunctions in a weak magnetic field.
Our findings offer a clearer understanding of LL physics in TMDs and may help interpret experimental observations in disordered systems. The connection we highlight between symmetry, topology, and disorder effects can be useful for future characterization of 2D materials.

\section{Acknowledgement}  
We thank 
Valentin Crépel, Daniel Muñoz-Segovia, and Julian Ingham for helpful discussions. This work is supported by the National Science Foundation under Award No. DMR-2340394 and the Sloan Foundation.

\section*{Appendix A: \texorpdfstring{$\vec{k}\cdot\vec{p}$}{k\,·\,p} Expansion and Orbital Magnetization}\label{app:kdotp}

In this Appendix, we provide a complementary derivation of the Landau level (LL) dispersion and orbital magnetization using the standard $\vec{k} \cdot \vec{p}$ expansion formalism \cite{wang2015magneto, xuan2020valley}. This approach offers additional physical insight by clearly separating the contributions from the effective mass and the orbital magnetic moment. We use this formalism to first demonstrate the fundamental limitations of two-band models before applying it to our three-band model and connecting it back to the results of the main text.

\subsection*{Landau level dispersion}
The starting point is the single-particle Hamiltonian in a magnetic field, $H=\frac{(\vec{p}+e \vec{A})^2}{2 m_e}+V(\vec{r})$, where $V(\vec{r})$ is the periodic lattice potential. To find the LL energies around a band extremum $\vec{K}$, one replaces the momentum deviation $\vec{k}$ with the operator $\hat{\vec{q}} = (\vec{p} + e\vec{A}) / \hbar$. To second order in perturbation theory, the effective Hamiltonian for states in band $\gamma$ is
\begin{equation}\label{eq:two_part_hamiltonian}
E_{\gamma}(\vec{K}+\hat{\vec{q}})=E_{\gamma \vec{K}}+\frac{\hbar^2\left(\hat{q}_x^2+\hat{q}_y^2\right)}{2 M_\gamma^*}-\vec{m}_{\gamma \vec{K}} \cdot \vec{B},
\end{equation}
where the effective mass $M_\gamma^*$ and orbital magnetic moment $\vec{m}_{\gamma \vec{K}}$ contain contributions from inter-band matrix elements:
\begin{align}
\frac{1}{M_\gamma^*}&=\frac{1}{m_e}+\frac{2}{m_e^2} \sum_{\gamma' \neq \gamma} \frac{\left|\braOket{u_{\gamma\vec{K}}}{\hat{p}_x}{u_{\gamma'\vec{K}}}\right|^2}{E_{\gamma \vec{K}}-E_{\gamma' \vec{K}}}, \label{eq:effective_mass}\\
\vec{m}_{\gamma \vec{K}}&=-i \frac{\mu_B}{\hbar} \sum_{\gamma' \neq \gamma} \frac{\braOket{u_{\gamma\vec{K}}}{\vec{p}}{u_{\gamma'\vec{K}}} \times \braOket{u_{\gamma'\vec{K}}}{\vec{p}}{u_{\gamma\vec{K}}}}{E_{\gamma \vec{K}}-E_{\gamma' \vec{K}}}.\label{eq:LK_mag_moment}
\end{align}
The orbital $g$-factor is defined by $g_{\rm orb, \gamma} = m_\gamma / \mu_B$ where $\mu_B = e\hbar / 2m_e$ is the Bohr magneton and $m_\gamma$ denotes the $z$ component (along the magnetic field) of the orbital magnetization.
The kinetic term gives rise to the standard LL ladder, $\epsilon_n^d = (n+1/2)\hbar\omega_c$ with $\omega_c=eB/M_\gamma^*$, while the second term is a Zeeman-like shift, $\epsilon^m = -m_{\gamma\vec{K}} B$. The total energy for the $n$-th LL is therefore
\begin{equation}\label{eq:two_contribution}
\epsilon_\gamma(n, B) = E_{\gamma\vec{K}} + \epsilon_n^d + \epsilon^m.
\end{equation}

\begin{figure}[htbp]
		\centering
		\includegraphics[width=0.48\textwidth] {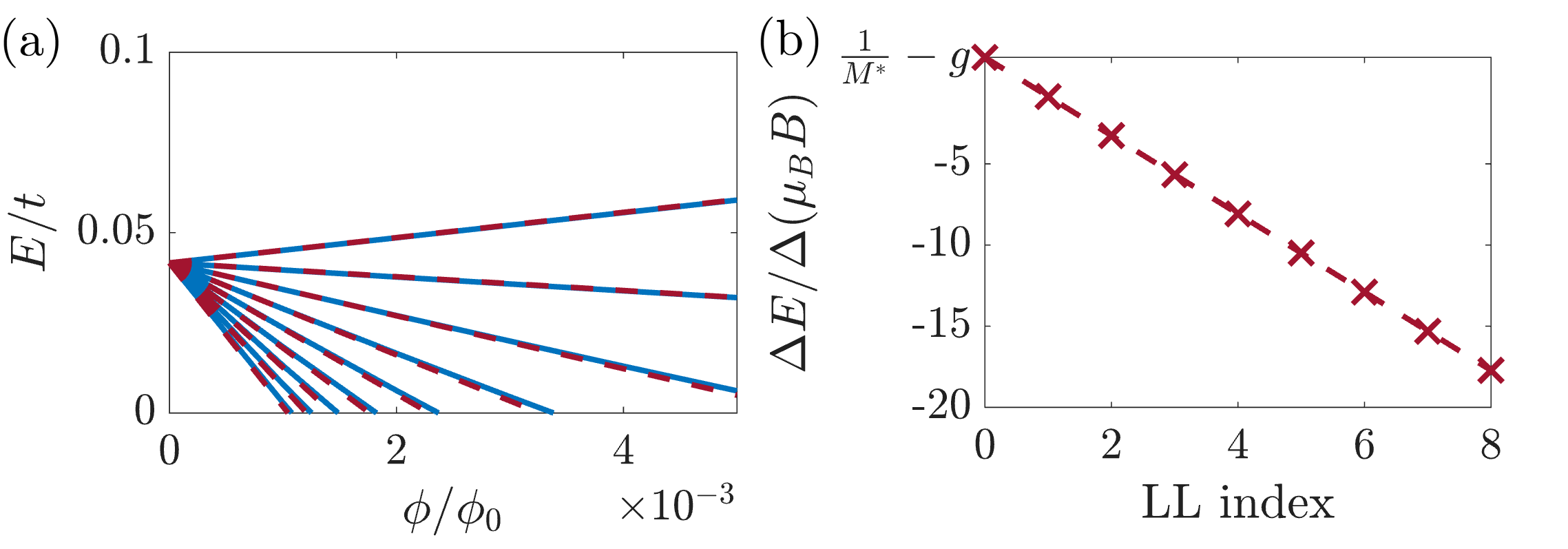}
		\caption{(a) The LL spectra of MoTe$_2$ around the valence band top as a function of magnetic flux in one unit cell in unit of the flux quantum $\phi_0 = h / e$. The blue lines indicate the results of perturbation calculation (Eq.~\eqref{eq:two_contribution}), and the red dashed line denotes LL obtained from minimal substitution (Eq.~\eqref{eq:C3_in_B}), which show perfect agreement. $t = 1$eV is the energy unit used. (b) The slope of the LLs in (a) as a function of LL index. The interception is given by $1/M^* - g$ and the slope of this line is $2 / M^*$.}
		\label{fig:LL_perturbation}
\end{figure}

As an example, we plot the LL spectra of MoTe$_2$ around the valence band top as a function of magnetic flux in Fig.~\ref{fig:LL_perturbation}. One can plot the LL slopes against the LL indices as in (b), the effective mass can be extracted from the slope which is $2 / M^*$, and the $g$-factor can be extracted from the interception which is $1 / M^* - g$.

\subsection*{Absence of magnetization asymmetry in a two band model}
The formalism above immediately reveals why a generic two-band model is insufficient to describe the LL spectrum in TMDs. For a two-band system consisting of a conduction band ($c$) and a valence band ($v$), the sum in Eq.~\eqref{eq:LK_mag_moment} contains only a single term. This directly leads to the constraint
\begin{equation}
\vec{m}_{c\vec{K}} = \vec{m}_{v\vec{K}}.
\end{equation}  
This enforces a perfectly symmetric orbital magnetization for electrons and holes. The experimentally observed asymmetry is therefore fundamentally incompatible with a two-band description and necessitates the inclusion of at least a third band. This conclusion remains valid even when spin is included, as long as it remains a good quantum number, since each spin sector must independently obey the two-band sum rule.

\subsection*{Origin of magnetization asymmetry in three band model}
In contrast, the $C_3$-symmetric three-band model naturally accommodates an asymmetric orbital magnetization. The cyclic structure of the Hamiltonian (Eq.~\eqref{eq:C3_Hk}) constrains the momentum matrix elements. In the $C_3$ eigenbasis, the momentum operators $\pi_{\pm} = \pi_x \pm i\pi_y$ become cyclic:
\begin{equation}
    \pi_+ = 
\left[\begin{array}{ccc}
0 & 0 & v_{02} \\
v_{10} & 0 & 0 \\
0 & v_{21} & 0
\end{array}\right],\ 
\pi_- = \pi_+^\dagger.
\end{equation}
Substituting this into the general expression for the magnetic moment simplifies it greatly, yielding
\begin{equation}
    m^z_{\gamma} = \mu_{\gamma, \gamma - 1} - \mu_{\gamma, \gamma + 1},
\end{equation}
where we have used the shorthand $\mu_{\gamma, \gamma'} \equiv |v_{\gamma, \gamma'}|^2 / (\epsilon_\gamma - \epsilon_{\gamma'})$ and a sign convention where a positive moment decreases the energy. A similar calculation for the effective mass gives
\begin{equation}
    \frac{\hbar e}{M^*_\gamma} = 2(\alpha_\gamma + \mu_{\gamma, \gamma + 1} + \mu_{\gamma, \gamma - 1}).
\end{equation}
Finally, we connect these results back to the perturbation theory in the main text. The slope of the zeroth LL ($n=0$) is given by $s_0^\gamma = \frac{\hbar e}{2M^*_\gamma} + m_\gamma$. Using the expressions above, this becomes
\begin{equation}
\begin{split}
    s_0^\gamma &= (\alpha_\gamma + \mu_{\gamma, \gamma+1} + \mu_{\gamma, \gamma-1}) + (\mu_{\gamma, \gamma-1} - \mu_{\gamma, \gamma+1})\\
    &= \alpha_\gamma + 2\mu_{\gamma, \gamma-1},
\end{split}
\end{equation}
which matches Eq.~\eqref{eq:LL_slope} exactly after accounting for a different index convention in the main text's formula.

\section*{Appendix B: Cyclicity of the LL Hamiltonian}\label{app:cyclicity}

In this appendix, we provide a more formal definition of the \emph{cyclicity} that constrains the Hamiltonian and gives rise to the valley-dependent selection rules for the Landau level (LL) wavefunctions.

\subsection*{Definition of Cyclicity}
The continuum Hamiltonian $H(\vec{k})$ (Eq.~\eqref{eq:C3_Hk}) belongs to a family of models parameterized by on-site energies $\epsilon_i$, curvatures $\alpha_i$, and hoppings $v_{ij}$. The action of the cyclic shift operator $X$ (Eq.~\eqref{eq:clock_shift}) on the orbital basis is equivalent to a cyclic permutation of the orbital indices ($0 \to 1 \to 2 \to 0$). While the Hamiltonian is not invariant under this operation (i.e., $[H, X] \neq 0$), its structure is preserved. Applying the operator maps the Hamiltonian to another one within the same family, but with cyclically permuted parameters:
\begin{equation}
    XH(\vec{k}; \epsilon_i, \alpha_i, v_{ij})X^{-1} = H(\vec{k}; \epsilon_{i-1}, \alpha_{i-1}, v_{i-1,j-1}),
\end{equation}
where all indices are understood modulo 3. We define this property, that the family of Hamiltonians is closed under the action of the shift operator, as \emph{cyclicity}. It is a generalization of a true symmetry, and it applies to both the continuum and lattice versions of the Hamiltonian.

\subsection*{Cyclicity and Valley Handedness}
The cyclicity of the Hamiltonian directly leads to selection rules for the LL wavefunctions. To quantify this, we define the composite operators $\hat{X}_{\circlearrowleft} = \hat{a} X$ and $\hat{X}_{\circlearrowright} = \hat{a} X^\dagger$, which combine an orbital shift with a lowering of the LL index. The expectation values of these operators act as order parameters for the cyclic structure of the wavefunctions.

Crucially, they distinguish the \emph{handedness} of the two valleys. For a magnetic field in the positive $z$-direction, the perturbed LL wavefunctions (Eq.~\eqref{eq:LL_perturbation_wf}) yield valley-contrasting expectation values:
\[
\begin{aligned}
&\text{$K$ valley:} && \langle\hat{X}_{\circlearrowright}\rangle \neq 0, \quad \langle\hat{X}_{\circlearrowleft}\rangle = 0,\\
&\text{$K'$ valley:} && \langle\hat{X}_{\circlearrowleft}\rangle \neq 0, \quad \langle\hat{X}_{\circlearrowright}\rangle = 0.
\end{aligned}
\]
Thus, the cyclic operators not only confirm the cyclic structure inherited from $C_3$ symmetry but also capture the valley-dependent chirality of the LL wavefunctions, which is directly responsible for optical selection rules.

\section*{Appendix C: Model Hamiltonians and Continuum Parameters} \label{app:model}

In this appendix, we provide the explicit forms of the tight-binding models used in the main text and detail their expansion around the K-point to obtain the parameters for the continuum model (Eq.~\eqref{eq:C3_Hk}).

\subsection*{Three-band TMD model}
We adopt the three-band tight-binding model for TMDs from Ref.~\cite{Liu2013}, which includes nearest-neighbor hoppings. In the basis of $(d_{z^2}, d_{xy}, d_{x^2 - y^2})$ orbitals, the momentum-space Hamiltonian is
\begin{equation}\label{eq:H_TMD}
H^{\operatorname{TMD}}(\vec{k})=\left[\begin{array}{ccc}
h_0 & h_1 & h_2 \\
h_1^* & h_{11} & h_{12} \\
h_2^* & h_{12}^* & h_{22}
\end{array}\right]
\end{equation}
where the matrix elements $h_{ij}$ are functions of momentum
\begin{align*}
h_0&=2 t_0(\cos 2 \alpha+2 \cos \alpha \cos \beta)+\epsilon'_1, \\
h_1&=-2 \sqrt{3} t_2 \sin \alpha \sin \beta+2 i t_1(\sin 2 \alpha+\sin \alpha \cos \beta), \\
h_2&=2 t_2(\cos 2 \alpha-\cos \alpha \cos \beta)+2 \sqrt{3} i t_1 \cos \alpha \sin \beta, \\
h_{11}&=2 t_{11} \cos 2 \alpha+\left(t_{11}+3 t_{22}\right) \cos \alpha \cos \beta+\epsilon'_2, \\
h_{22}&=2 t_{22} \cos 2 \alpha+\left(3 t_{11}+t_{22}\right) \cos \alpha \cos \beta+\epsilon'_2, \\
h_{12}&=\sqrt{3}\left(t_{22}-t_{11}\right) \sin \alpha \sin \beta +4 i t_{12} \sin \alpha(\cos \alpha-\cos \beta)
\end{align*}
with $(\alpha, \beta) \equiv \left( k_x a/2, \sqrt{3} k_y a/2\right)$.
The hopping parameters are obtained from the paper \cite{Liu2013} with generalized gradient approximation.
and tight-binding parameters given in Ref.~\cite{Liu2013}.

To connect this lattice model to our continuum theory, we expand the Hamiltonian around the $K = (4\pi/3, 0)$ point and perform a basis transformation to the $C_3$ eigenbasis $(|0\rangle, |1\rangle, |2\rangle)$. This procedure yields the continuum parameters used in Eq.~\eqref{eq:C3_Hk}:
\begin{equation}
\begin{aligned}
\epsilon_{0} &= \epsilon'_1 - 3 t_0\\ \epsilon_{1} &= \epsilon'_2 - \frac{3}{2}(t_{11} + t_{22} + 2 \sqrt{3} t_{12})\\ \epsilon_{2} &= \epsilon'_2 - \frac{3}{2}(t_{11} + t_{22} - 2 \sqrt{3} t_{12})\\
\alpha_{0} &= \frac{3}{4}t_0, \\
\alpha_{1} &= \frac{3}{8}(t_{11} + t_{22} + 2 \sqrt{3} t_{12}) \\
\alpha_{2} &= \frac{3}{8}(t_{11} + t_{22} - 2 \sqrt{3} t_{12})\\
v_{10} &= \frac{3\sqrt{2}i}{4}(t_1 + \sqrt{3}t_2)\\
v_{21} &= -\frac{3\sqrt{3}}{4}(t_{11} - t_{22})\\
v_{02} &= -\frac{3\sqrt{2}i}{4}(t_1 - \sqrt{3}t_2).    
\end{aligned}
\end{equation}
The numerical values for these parameters for MoTe$_2$ and WS$_2$ are provided in Table~\ref{tab:continuum_params}.

\subsection*{Distorted Kagome Lattice Model}
The tight-binding Hamiltonian for the distorted kagome lattice (Eq.~\eqref{eq:kagome_TB}) in momentum space is given by
\begin{align}
&H^{\text{kagome}}(\vec{k})=\nonumber\\
-
&\left[\begin{array}{ccc}
0 & t + t^\prime e^{-i\vec{k}\cdot\vec{a}_1} & t + t^\prime e^{-i\vec{k}\cdot\vec{a}_2} \\
t + t^\prime e^{i\vec{k}\cdot\vec{a}_1} & 0 & t + t^\prime e^{-i\vec{k}\cdot(\vec{a}_1 - \vec{a}_2)} \\
t + t^\prime e^{i\vec{k}\cdot\vec{a}_2} & t + t^\prime e^{i\vec{k}\cdot(\vec{a}_1 - \vec{a}_2)} & 0
\end{array}\right]
\end{align}
where $\vec{a}_1 = (1, 0)$ and $\vec{a}_2 = (1/2, \sqrt{3}/2)$ are the lattice vectors (lattice constant set to unity). The parameters $t$ and $t^\prime$ are the nearest-neighbor intra- and inter-cell hopping strengths, respectively. We have chosen a gauge where all atoms in the unit cell are considered to be at the same position. 

Expanding this Hamiltonian around the $K$ point yields the same functional form as Eq.~\eqref{eq:C3_Hk}. For the chosen gauge where all sublattices are at the origin, the coupling $v_{02}$ vanishes. The continuum parameters are: 
\begin{equation}
\begin{aligned}
    &\epsilon_{0} = t^\prime - 2t,\quad \epsilon_{1} = -2t^\prime + t,\quad \epsilon_{2} = t + t^\prime,\\ 
    &\alpha_{0} = -\frac{1}{4}t^\prime, \quad \alpha_{1} = \frac{1}{2}t^\prime,\quad \alpha_{2} = -\frac{1}{4}t^\prime,\\ 
    &v_{10} = -\frac{\sqrt{3} - 3i}{4}t^\prime,\quad v_{21} = \frac{\sqrt{3} - 3i}{4}t^\prime,\quad v_{02} = 0.
\end{aligned}
\end{equation}

\begin{table}[h]
\caption{Parameters of the $C_3$-Symmetric Continuum Model at the K-point, derived from the tight-binding model of Ref.~\cite{Liu2013}. These parameters are used to analytically calculate quantities such as the Landau level slope $s_\gamma^0$. Everything is in units of eV.}
\label{tab:continuum_params}
\begin{ruledtabular}
\begin{tabular}{lccc}
Parameter & MoTe$_2$ Value & WS$_2$ Value \\
\hline
$\epsilon_0$ & 1.1120 & 1.564  \\
$\epsilon_1$ & 0.0416 & 0.024  \\
$\epsilon_2$ & 2.5254 & 3.443  \\
$|v_{10}| (=|v_{01}|)$ & 0.9583 & 1.3776  \\
$|v_{21}| (=|v_{12}|)$ & 0.0585 & 0.2975  \\
$|v_{02}| (=|v_{20}|)$ & 0.4746 & 0.4081 \\
$\alpha_0$ & -0.1268 & -0.1552  \\
$\alpha_1$ & 0.4826 & 0.5388  \\
$\alpha_2$ & -0.1383 & -0.3160  \\
\end{tabular}
\end{ruledtabular}
\end{table}

\section*{Appendix D: Complete Hofstadter butterfly spectra} \label{app:butterfly}

\begin{figure}[htbp]
    \centering
    \includegraphics[width=0.48\textwidth] {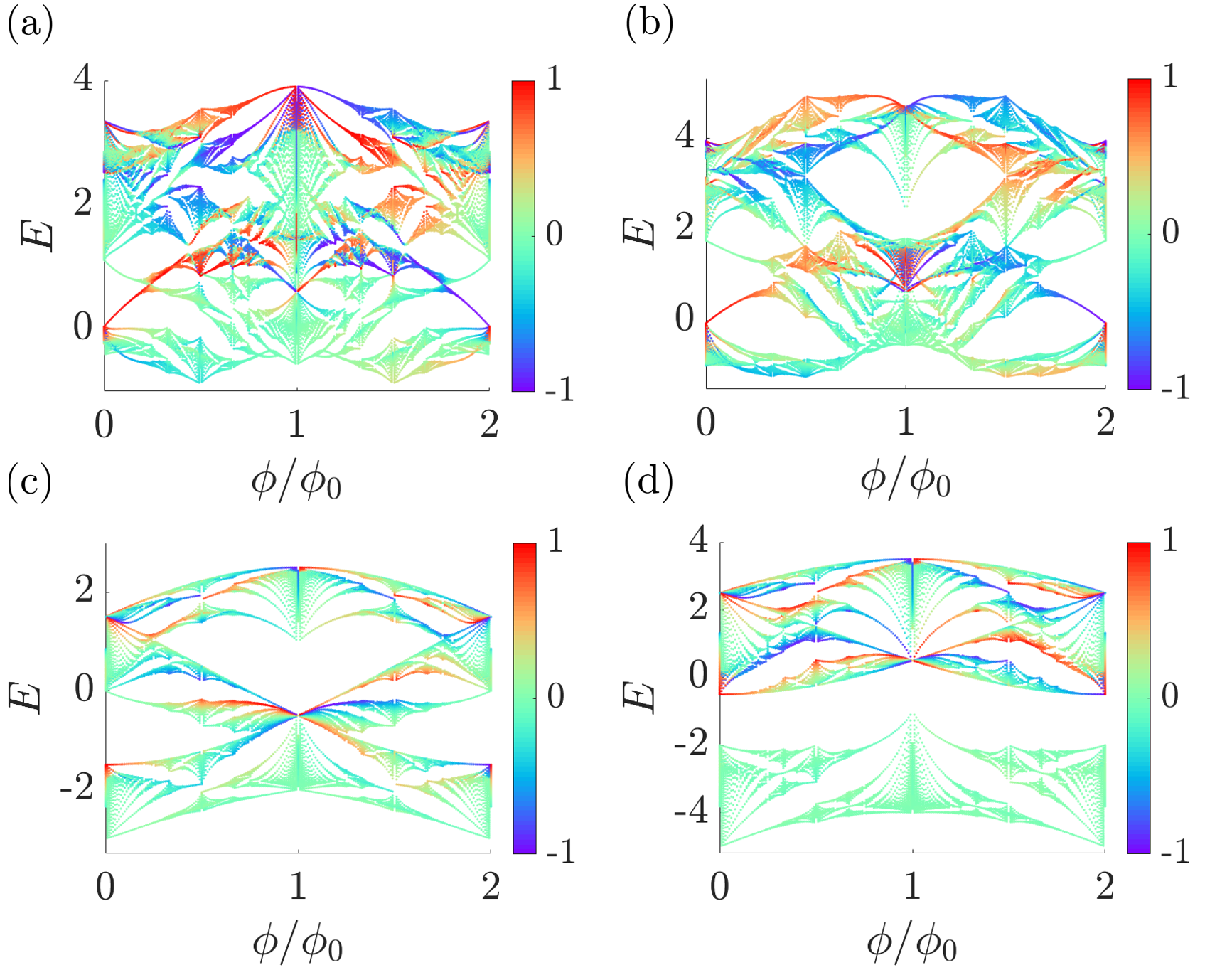}
    \caption{The Hofstadter butterfly spectrum of the three-band model for (a) MoTe$_2$, (b) WSe$_2$ and the distorted kagome lattice model with (c) $t = 0.5, t^\prime = 1$ and (d) $t = 1.5, t^\prime = 1$. For the TMDs, the energy unit is set to $1$eV.
    $\phi_0 = h / e$ is the flux quantum. The color denotes the expectation value of $S = \frac{3}{2\pi} \operatorname{arg} Z$.}
    \label{fig:whole_butterfly}
\end{figure}

In this Appendix, we provide the whole Hofstadter butterfly spectrum \cite{hofstadter1976energy} for the three-band model of TMDs and the distorted kagome lattice model, which complements our results in Sec.\ref{sec:butterfly}. The results are shown in Fig.~\ref{fig:whole_butterfly}. From the comparison between (a) and (b), we see that the anomalous dispersion of the zeroth LL is not specific to MoTe$_2$, but is shared by other TMD materials as well. Also, (a)-(c) shares the same connectivity, i.e., the Hofstadter bands coming from the valence band connect to those from the conduction band in energy, a feature consistent with their non-trivial OAL topology. However, in (d), the valence band spectra is always separated from the conduction band spectra by a finite energy gap, characteristic of a topologically trivial band structure.

\section*{Appendix E: Disorder immunity of Graphene LLs} \label{app:graphene}

In this Appendix, we provide the details of the continuum and lattice models for gapped graphene discussed in Sec.~\ref{sec:disorder}.
The low-energy physics in the $K$ valley is captured by the Dirac Hamiltonian (the $K'$ valley being related by transposition)
\begin{equation}
    H = 
    \begin{bmatrix}
        m & v \hat a \\
        v \hat a^\dagger & -m
    \end{bmatrix},
\end{equation}
where \( \hat a = -i(\partial_x + i\partial_y) - e(A_x - iA_y) \), \( v = \sqrt{3}t / 2\hbar \) denotes the Fermi velocity, and \( \vec{A} \) is the vector 
potential. 
The zeroth Landau level (LL) has the wavefunction \( |\psi_0\rangle = (0, |0\rangle)^T \), with \( \hat a|0\rangle = 0 \). 
Perturbations in the off-diagonal channels, of the form
\begin{equation}
    \delta H = \delta v 
    \begin{bmatrix}
        0 & \hat a \\
        \hat a^\dagger & 0
    \end{bmatrix},
\end{equation}
do not affect the zeroth LL, since \( \delta H |\psi_0\rangle = 0 \). 
In contrast, diagonal perturbations act uniformly on all LLs, leading to a broadening of the spectrum without selective protection of the zeroth level.

We numerically simulate the disorder effect in gapped graphene system using a real-space tight-binding model
\begin{equation} \label{eq:TB_graphene}
\begin{split}
    H = -&\sum_{\langle ij\rangle} (t + \delta t_{ij}) e^{i(\phi_{ij} + \theta_{ij})} c_i^\dagger c_j 
    + \sum_i \varepsilon_i c_i^\dagger c_i \\
    &+ m \sum_i (c_{iA}^\dagger c_{iA} - c_{iB}^\dagger c_{iB}),
\end{split}
\end{equation}
where \( \langle ij \rangle \) denotes nearest neighbors, \( \phi_{ij} \) is the Peierls phase from a uniform magnetic flux, and \( \delta t_{ij} \), \( 
\theta_{ij} \), and \( \varepsilon_i \) introduce disorder in hopping amplitude, phase, and on-site potential, respectively. The mass term \( m \) opens a 
gap between sublattices. We simulate the model on an \( N \times N \) honeycomb lattice with periodic boundary conditions to eliminate edge effects.\ We set 
the lattice constant to be unity.

To model spatially correlated disorder, we follow Refs.~\cite{PhysRevB.75.235317, Tohru2009quantum}, generating disorder fields with Gaussian distributions:
\begin{equation} \label{eq:disorder_strength_distribution}
    P(\varepsilon) = \frac{1}{\sqrt{2\pi\sigma^2}} e^{-\varepsilon^2/2\sigma^2}, \quad 
\langle \varepsilon_i \varepsilon_j \rangle = \sigma^2 e^{-|\vec{R}_i - \vec{R}_j|^2 / 4\eta^2},
\end{equation}
with correlation length \( \eta \). The disorder fields are constructed numerically via Gaussian filtering. Hopping and phase disorders are generated 
analogously.

\newpage

\bibliography{ref_new.bib}

\end{document}